\documentclass[sigconf,authorversion]{acmart}

\usepackage{multirow}
\usepackage{graphicx}
\usepackage{float}

\usepackage{xcolor}  %% temporary for highlighting text that requires paraphrasing

\renewenvironment{quote}{\list{}{\rightmargin\leftmargin}\item\relax
}{\endlist}

\renewenvironment{quote}
  {\list{}{\rightmargin0.0in\leftmargin0.2in}\item\relax}
  {\endlist}

\AtBeginDocument{%
  \providecommand\BibTeX{{%
    \normalfont B\kern-0.5em{\scshape i\kern-0.25em b}\kern-0.8em\TeX}}}

\copyrightyear{2024}
\acmYear{2024}
\setcopyright{rightsretained}
\acmConference[ITiCSE 2024]{Proceedings of the 2024 Innovation and Technology in Computer Science Education V. 1}{July 8--10, 2024}{Milan, Italy}
\acmBooktitle{Proceedings of the 2024 Innovation and Technology in Computer Science Education V. 1 (ITiCSE 2024), July 8--10, 2024, Milan, Italy}
\acmDOI{10.1145/3649217.3653574}
\acmISBN{979-8-4007-0600-4/24/07}

\makeatletter
\gdef\@copyrightpermission{
  \begin{minipage}{0.3\columnwidth}
   \href{https://creativecommons.org/licenses/by/4.0/}{\includegraphics[width=0.90\textwidth]{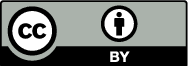}}
  \end{minipage}\hfill
  \begin{minipage}{0.7\columnwidth}
   \href{https://creativecommons.org/licenses/by/4.0/}{This work is licensed under a Creative Commons Attribution International 4.0 License.}
  \end{minipage}
  \vspace{5pt}
}
\makeatother

\begin{document}

\title{Desirable Characteristics for AI Teaching Assistants in Programming Education}
%% \title{Alignment of Student Perceptions of Digital TA Characteristics with Educational Objectives}
%% \title{Student Reflections on Essential Characteristics for AI Assistants in Programming Education}

\author{Paul Denny}
\orcid{0000-0002-5150-9806}
\affiliation{
  \institution{University of Auckland}
  \city{Auckland}
  \country{New Zealand}
}
\email{paul@cs.auckland.ac.nz}

\author{Stephen	MacNeil}
\orcid{0000-0003-2781-6619}
\affiliation{
  \institution{Temple University}
  \city{Philadelphia}
  \state{PA}
  \country{United States}}
\email{stephen.macneil@temple.edu}

\author{Jaromir Savelka}
\orcid{0000-0002-3674-5456}
\affiliation{
  \institution{Carnegie Mellon University}
  \city{Pittsburgh}
  \state{PA}
  \country{United States}
}
\email{jsavelka@cs.cmu.edu}

\author{Leo Porter}
\orcid{0000-0003-1435-8401}
\affiliation{
  \institution{University of California, San Diego}
  \city{San Diego}
  \state{CA}
  \country{United States}
}
\email{leporter@ucsd.edu}

\author{Andrew Luxton-Reilly}
\orcid{0000-0001-8269-2909}
\affiliation{
  \institution{University of Auckland}
  \city{Auckland}
  \country{New Zealand}
}
\email{a.luxton-reilly@auckland.ac.nz}

%% Authors
% \author{Anonymous Author}
% \affiliation{
%   \institution{Anonymous Institution}
%   \city{Anonymous City}
%   \country{Country}
% }
% \email{anon@anon.anon}

% \author{Anonymous Author}
% \affiliation{
%   \institution{Anonymous Institution}
%   \city{Anonymous City}
%   \country{Country}
% }
% \email{anon@anon.anon}

% \author{Anonymous Author}
% \affiliation{
%   \institution{Anonymous Institution}
%   \city{Anonymous City}
%   \country{Country}
% }
% \email{anon@anon.anon}

% \author{Anonymous Author}
% \affiliation{
%   \institution{Anonymous Institution}
%   \city{Anonymous City}
%   \country{Country}
% }
% \email{anon@anon.anon}

% \author{Anonymous Author}
% \affiliation{
%   \institution{Anonymous Institution}
%   \city{Anonymous City}
%   \country{Country}
% }
% \email{anon@anon.anon}

% \author{Paul Denny}
% \orcid{0000-0002-5150-9806}
% \affiliation{
%   \institution{The University of Auckland}
%   \city{Auckland}
%   \country{New Zealand}
% }
% \email{paul@cs.auckland.ac.nz}

%\renewcommand{\shortauthors}{Paul Denny, et al.}

\renewcommand{\shortauthors}{Paul Denny, Stephen MacNeil, Jaromir Savelka, Leo Porter, \& Andrew Luxton-Reilly}

\begin{abstract}
Providing timely and personalized feedback to large numbers of students is a long-standing challenge in programming courses.  Relying on human teaching assistants (TAs) has been extensively studied, revealing a number of potential shortcomings. These include inequitable access for students with low confidence when needing support, as well as situations where TAs provide direct solutions without helping students to develop their own problem-solving skills.  With the advent of powerful large language models (LLMs), digital teaching assistants configured for programming contexts have emerged as an appealing and scalable way to provide instant, equitable, round-the-clock support.  Although digital TAs can provide a variety of help for programming tasks, from high-level problem solving advice to direct solution generation, the effectiveness of such tools depends on their ability to promote meaningful learning experiences.  If students find the guardrails implemented in digital TAs too constraining, or if other expectations are not met, they may seek assistance in ways that do not help them learn. Thus, it is essential to identify the features that students believe make digital teaching assistants valuable.  We deployed an LLM-powered digital assistant in an introductory programming course and collected student feedback ($n=813$) on the characteristics of the tool they perceived to be most important.  Our results highlight that students value such tools for their ability to provide instant, engaging support, particularly during peak times such as before assessment deadlines.  They also expressed a strong preference for features that enable them to retain autonomy in their learning journey, such as scaffolding that helps to guide them through problem-solving steps rather than simply being shown direct solutions.

\end{abstract}




\maketitle

\section{Introduction}

High-quality and timely feedback is essential in educational settings, particularly for students in large undergraduate programming courses who often have a wide variety of prior experience. 
Human teaching assistants (TAs) have helped address this need,  but scaling their support to match classes that are growing in size and diversity remains a significant challenge.
%Human teaching assistants (TAs) have traditionally catered to this need but scaling their support to match classes that are growing in size and diversity remains a significant challenge.  
Prior literature on student interactions with TAs has revealed mismatches between student expectations and educational objectives.  In particular, students grappling with impending deadlines tend to prioritize assignment completion over long-term learning and conceptual mastery \cite{lim2023student}.  In practice, TAs frequently reveal solutions during help sessions instead of guiding students through the debugging process, further entrenching problem-solving dependencies rather than fostering independence and resilience \cite{krauselevy2022exploration}.  Another limitation of human TAs is their availability and accessibility.  Providing immediate help in programming courses is difficult, with scheduled office hours being underutilized \cite{smith2017office} and in-person support failing to reach all students equally, as some may hesitate to seek assistance while others monopolize it \cite{smith2017my}. 

The scalability and performance of large language models (LLMs) has prompted considerable interest in their use for powering digital teaching assistants, offering much higher availability and faster response times than human TAs.   However, a critical consideration for instructors is the ability of digital TAs to support healthy learning interactions.  There are concerns that students may become overly dependent on these tools for solutions. This calls for built-in ``guardrails'' to prevent such misuse \cite{denny2023computing}.  However, if the constraints imposed by such guardrails are overly prohibitive, or a digital TA lacks some other desired characteristics, then students may %refrain from using it and 
instead turn to publicly available unconstrained LLMs which may not be as pedagogically useful.  Presently, very little is known about the specific characteristics that students value in digital TAs, and how these might differ from their expectations of human tutors, despite the fact that such tools are likely to become ubiquitous. 

To address this gap, we deployed a digital TA in a large undergraduate introductory programming course, in order to assess students' perceptions and preferences.  As part of a structured lab activity, students observed LLM-generated responses to their own queries for debugging and code writing assistance on sample problems.  After the lab, students were also optionally able to use the digital TA to seek assistance for their own project work.  Students evaluated the quality of the feedback produced by the digital TA to their queries, and also reflected on the characteristics that they felt were most important in a digital TA.  This research focuses on two central questions regarding digital TA deployment:

\begin{enumerate}
    \item[\textbf{RQ1:}] How do students perceive the effectiveness of a digital TA's responses concerning debugging and code writing tasks in terms of correctness and helpfulness?
    \item[\textbf{RQ2:}] What attributes of a digital programming assistant do students consider most essential for supporting their learning in introductory programming courses?
\end{enumerate}

%\noindent In general, students found that the help provided by the digital TA was both highly accurate and helpful, and they particularly valued feedback that explained underlying principles and showed them how to obtain solutions rather than being provided with solutions directly.
%This paper contributes novel insights into the characteristics that students value in a digital TA, and can inform the design of future tools that will effectively support student learning. 

\section{Related work}

\subsection{LLMs in Computing Education}

%%%% Text is set to red as this requires rewriting (it is taken from a prior paper).  If anyone rewrites, please remove the red text to signal this!
%%%% Jaromir: I paraphrased and restructured this. Hence, I think it is ok now and the red text is removed.
LLMs have been extensively studied in the context of computing education \cite{prather2023robots}. There is a growing body of work focusing on the capabilities of LLMs for solving programming exercises~\cite{finnie2022robots,denny_conversing_2023,hou2023more,savelka2023can}, as well as answering \cite{savelka2023thrilled} and generating programming-related MCQs \cite{tran2023generating}. Other studied tasks include bug detection \cite{macneil2023decoding}, test suite authoring \cite{chen_codet_2022}, automatic generation of programming exercises \cite{sarsa_automatic_2022}, code explanations \cite{leinonen_comparing_2023, macneil2023experiences}, and worked examples \cite{jury2024evaluating}, and the ways that students interact with code-generating LLMs when programming \cite{kazemitabaar2023studying, prather_its_2023}. When employed as digital TAs, it is not clear if LLMs are as effective as human tutors \cite{savelka2023thrilled,nguyen_empirical_2022}.  For example, a study in the context of mathematics education by Pardos and Bhandari found that significant learning gains were only observed for students who were provided human-created hints, and not for those who were given LLM-generated hints 
\cite{pardos2023learning}. 
There remains much ongoing discussion and work exploring the quality and completeness of feedback generated by LLMs in computing education contexts \cite{hellas2023exploring, balse2023investigating}.

%Furthermore, there have been concerns around student over-reliance on AI-generated code \cite{prather_its_2023,becker2023programming,brusilovsky2023future,collins2023policy}. For example, Kazemitabaar et al. found that around half the time students submitted an AI-generated code without making any changes to it \cite{kazemitabaar2023studying}. 

%Finally, it does not appear that it is possible to provide students with unmoderated responses from LLMs as these may be inaccurate or unsuitable. Hellas et al. analyzed LLM generated responses to 150 student help requests, finding not all of the issues in the students' help requests were found and false positives were common \cite{hellas2023exploring}.  Balse et al. reported high variability in the accuracy of LLM-generated feedback on student submissions \cite{balse2023investigating}. Thus, there is an important need to carefully study the use of LLM-powered tutors in the context of computing classrooms.

%\subsection{Student expectations of TA interactions}
\subsection{Student and TA interactions in CS}

%%%% Anyone can tackle this, but Leo this is based on Leo's prior work.  It would be good to connect it more strongly to the current work.
Prior work has shown that student and TA interactions are challenging for both students and tutors. \citeauthor{krauselevy2022exploration} examined interactions between students and undergraduate TAs during help-seeking sessions focused on debugging assignments in introductory computing courses~\cite{krauselevy2022exploration}.  They found that the TAs often gave students the solution without teaching them the underlying concepts of the process of debugging. This behavior results in students leaving tutoring sessions with resolved issues but potentially without a deeper understanding of how to independently tackle similar problems in the future.

Human TAs face challenges providing effective help for students because of a need to triage and prioritize multiple student requests, engage students effectively, and balance teaching with time constraints~\cite{markel2021inside,riese2021challenges}. In addition, TAs must deal with their own understanding of the material, unprepared students, and diagnosing student knowledge to provide appropriate assistance~\cite{riese2021challenges}.

Most directly related to this work is that of \citet{lim2023student} who explored student expectations of human TAs.  The main findings from interviews with students in a CS2 course indicate that student expectations and desires can be misaligned with their learning goals, particularly under the pressure of imminent assignment deadlines. While students acknowledge the importance of understanding and learning from tutors, many prefer direct answers or solutions as deadlines approach to prioritize completing assignments over deeper learning. Additionally, students do not always value the importance of identifying their own bugs and may expect tutors to do it for them, potentially creating a reliance on tutors and compromising long-term learning.

\section{Methods}

\subsection{Course context and LLM}

Our data was collected from a large introductory programming course taught at the University of Auckland, a large public research university in New Zealand, in 2023.  A total of 889 students were enrolled in the course which spans a 12-week teaching term.  Every week, students complete a set of laboratory tasks 
%(primarily, auto-graded programming problems) which 
each contributing 2\% towards their final grade in the course.  

% \begin{figure}
% \centering
%   \includegraphics[width=.9\columnwidth]{figs/Cost.png}
%   \caption{The daily cost of running the study (9--21 October 2023) using the GPT-4 model provided by OpenAI.  The total across over the period was US\$509.51.}
%   \label{fig:cost}
% \end{figure}

For the purposes of the current study, we provided students access to an LLM-powered digital assistant designed for programming courses called CodeHelp \cite{liffiton2024codehelp}.  CodeHelp is open source\footnote{\url{https://github.com/liffiton/Gen-Ed}} and is one of several applications in a publicly available framework that can be used 
%by educators and researchers 
for building web applications that use LLMs for educational purposes.  Although CodeHelp is open source, an OpenAI API key must be provided so that it can make requests to one of OpenAI's LLMs.  For this study we chose to use GPT-4, which was the most capable model at the time but also the most expensive.  The total cost for running the study over 13 days (9--21 October 2023) was US\$510.

%Figure \ref{fig:cost} shows the daily API costs associated with running this study. 

CodeHelp is designed to answer free-form student inquiries while maintaining a key safeguard: it does not generate solution code, irrespective of the query's phrasing.  This is achieved with a series of distinct prompts which include explicit instructions not to generate code, and which separately verify that code is absent in any responses received from the LLM.  This design is intended to allow students to receive step-by-step guidance in a mixture of natural language and pseudocode, as well as explanations of ideas to develop conceptual understanding, but require that they construct syntactically correct code on their own.  CodeHelp also uses prompts to generate responses that are ``positive and encouraging'', much like a human tutor would be expected to.  Examples of this tone can be seen in the last line of the outputs shown in Figures \ref{fig:example1} and \ref{fig:example2}.

%CodeHelp uses a dual-prompt approach, where the first prompt assesses the clarity and completeness of a student's input, and responds with a request for additional information if necessary. The second prompt creates an initial `main' response using the student's query, combined with explicit directives to avoid providing solution code. If the `main' response mistakenly contains code, a third prompt is employed to remove the code from the response.

\subsection{Lab tasks}

\begin{figure}
\centering
  \includegraphics[width=.8\columnwidth]{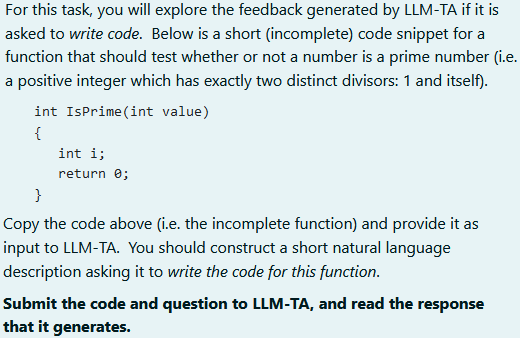}
  \caption{Task 1 -- writing code for the IsPrime() function}
  \label{fig:screen1}
\end{figure}

\begin{figure}
\centering
  \includegraphics[width=.8\columnwidth]{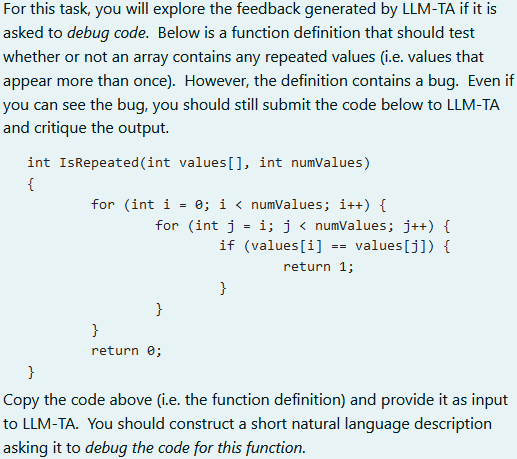}
  \caption{Task 2 -- debugging the IsRepeated() function}
  \label{fig:screen2}
\end{figure}

\begin{figure}
\centering
  \includegraphics[width=.8\columnwidth]{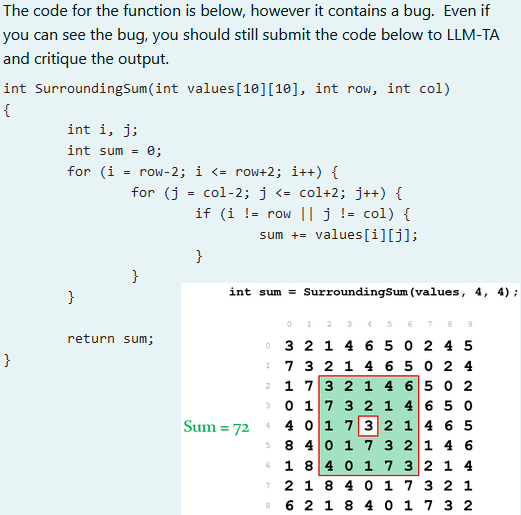}
  \caption{Task 3 -- debugging the SurroundingSum() function}
  \label{fig:screen3}
\end{figure}

To ensure that all students had a comparable experience with CodeHelp, we prepared three tasks that were included on the lab (Week 11). These tasks consisted of one \emph{code writing} and two \emph{debugging} problems.  For each task students were provided some sample code and were asked to construct an appropriate query for CodeHelp.  Each task was followed by two evaluation questions where students rated the correctness and helpfulness of the feedback produced by CodeHelp.  Figures \ref{fig:screen1}, \ref{fig:screen2} and \ref{fig:screen3} show the three tasks. 

%\subsubsection{Code writing assistance}

\noindent \textbf{Code writing assistance:}
Task 1 involved \emph{code writing}, and required students to provide the code fragment (for the incomplete \texttt{IsPrime} function in Figure \ref{fig:screen1}) to CodeHelp and construct a query asking it to complete the code.  Submissions for Task 1 tended to be relatively short, as exemplified by the following queries:

\begin{itemize}
    \item \emph{Could you please write me a function that determines whether an integer is a prime number or not?}
%    \item \emph{The function is meant to return a prime number. Write the code to complete this.}
    \item \emph{Can you write the code for this function?}
\end{itemize}

One example response to the query \emph{``Can you write the code for this function?''}, when the code from Figure \ref{fig:screen1} was included verbatim, is shown in Figure \ref{fig:example1}.

\begin{figure}[ht]
  \centering
  \footnotesize
  \framebox{
    \begin{minipage}{\columnwidth} % You can adjust the width to what you need.
Sure, I can help you understand how to write a function to check if a number is prime in C.  A prime number is a natural number greater than 1 that has no positive divisors other than 1 and itself. The first few prime numbers are 2, 3, 5, 7, 11, and so on.  To check if a number is prime, you can follow these steps:

1. If the number is less than 2, it's not prime. So, you can return 0.

  \textbf{ \emph{<Steps 2 and 3 not shown for space reasons>}}
%2. If the number is 2, it's prime. So, you can return 1.

%3. For any number greater than 2, you can check divisibility from 2 to the square root of the number. If any of these numbers can divide the input number exactly, then the number is not prime. You can return 0.

4. If the number is not divisible by any number in the above step, then the number is prime. You can return 1.

Remember to use a loop for step 3. The reason we only need to check divisibility up to the square root of the number is because a larger factor of the number would be a multiple of a smaller factor that has already been checked.

I hope this helps! Let me know if you have any other questions.
    \end{minipage}
  }
  \caption{Example response from CodeHelp to Task 1.}
  \label{fig:example1}
\end{figure}

%\subsubsection{Debugging assistance}

\noindent \textbf{Debugging assistance:}
The other two tasks were \emph{debugging} tasks, where students were asked to query CodeHelp to identify an error in buggy code. 
%prompted to provide a complete (but buggy) code fragment to CodeHelp and construct a query to ask it to debug the code.  
Examples of queries for Tasks 2 and 3 included:

\begin{itemize}
%    \item \emph{I need to create a function that tests whether an array contains any repeated values. Can you help me debug the code I have written?}
    \item \emph{I want to use this code to determine if there is a repeated value in the input array, but there seems to be a bug}
    \item \emph{it contains a bug}
\end{itemize}

An example response to the query \emph{``it contains a bug''}, when the code for Task 2 (see Figure \ref{fig:screen2}) was included verbatim, is the output shown in Figure \ref{fig:example2}.

\begin{figure}[ht]
  \centering
  \footnotesize
  \framebox{
    \begin{minipage}{\columnwidth} % You can adjust the width to what you need.
The issue with your function lies in the nested for loop. You're comparing each element in the array with itself and all the subsequent elements. This means that each element will always find itself as a duplicate because you're starting the inner loop from `i` instead of `i + 1`.

In your current implementation, when `i` and `j` are the same, `values[i]` is equal to `values[j]`, and your function will return `1` indicating that a duplicate is found, even if there are no duplicates in the array.

%  \textbf{ \emph{<Suggested fix, starting inner loop from i+1, not shown for space reasons>}}

To fix this, you should start the inner loop from `i + 1` instead of `i`. This way, you're comparing each element with all the elements that come after it, not including itself.

Remember, when dealing with nested loops, especially in cases where you're comparing elements in an array, it's important to carefully consider where each loop starts and ends to avoid comparing an element with itself. 

Keep going, you're doing great!
    \end{minipage}
  }
  \caption{Example response from CodeHelp to Task 2.}
  \label{fig:example2}
\end{figure}

%\subsubsection{Evaluations of correctness and helpfulness}

\noindent \textbf{Evaluations of correctness and helpfulness:}
After each of these tasks, students were asked to evaluate the quality of the response that was generated by CodeHelp, with respect to \emph{correctness} and \emph{helpfulness}.  Specifically, students responded to the following two Likert-scale items for each of the three programming tasks:

\begin{enumerate}
    \item The response from CodeHelp is \textbf{correct} (i.e. it does not include any false or inaccurate content)
    \item The response from CodeHelp is \textbf{helpful} (i.e. is a useful response for a student asking this question)
\end{enumerate}

\subsection{Reflections on desirable characteristics}

%three programming tasks and an open-response reflection question that prompted students to think about the important characteristics of a digital TA for use in programming courses.  

After evaluating the feedback generated by CodeHelp to the three programming tasks, students were asked to think about and describe the characteristics they felt were important for a digital TA used in programming courses.  Specifically, they were asked to answer the following: %prompted with the following statement: 
\emph{Now that you have observed feedback from CodeHelp in response to requests for writing and debugging code, answer the following reflection question:} \emph{What do you think are the most important characteristics for an automated programming assistant (like CodeHelp) if it is to be suitable for use in a programming course?}

%\begin{quote}
 %   \item What do you think are the most important characteristics for an automated programming assistant (like CodeHelp) if it is to be suitable for use in a programming course?
%\end{quote}

The deadline for completing the lab tasks and for submitting the reflection was midnight on the evening of October 16th.  After this date, students were invited to use CodeHelp optionally during the week following the lab when they were working on their end-of-course projects (worth 12\% of their final grade).  %This project was an implementation of a variation of the classic ``battleships'' game.  
Students were permitted to use CodeHelp to ask any questions related to the final project, which was due at midnight on the evening of October 21st.  
%Given that use of CodeHelp after the lab was ungraded, we measure interactions during this time as an indication of the value they see in such a tool to support their coursework. 

\subsection{Data analysis}

We report student interactions with CodeHelp in two ways.  
First, we perform a quantitative analysis of the frequency with which students interact with the tool over the duration of the study (9--21 October) and we summarize student ratings of the correctness and accuracy of the responses generated by CodeHelp. %Firstly, we present a heatmap visualising when queries were made over the duration of the study (9--21 October).  In addition, we explore the quantity of interactions per student both before the lab deadline (where completing the three tasks was graded) and after the lab deadline (where use of CodeHelp for project support was ungraded). For student perceptions of the correctness and helpfulness of the responses generated by CodeHelp, we summarize these using a grouped bar chart (showing the number of responses at each level of agreement on the Likert scale). 

Second, our qualitative analysis of the open-response reflection data follows the guidelines for reflexive thematic analysis outlined by Braun and Clarke \cite{braun2006using}.  In this study, three of the authors independently coded a sample of responses with succinct labels, and subsequently met to discuss the codes and to develop higher-level themes\footnote{\url{https://www.thematicanalysis.net/doing-reflexive-ta/}}.  When presenting the results, we report the most common themes and illustrate these with examples of student responses.

\section{Results}

\subsection{Interactions with the tool}

\begin{figure}
\centering
  \includegraphics[width=0.9\columnwidth]{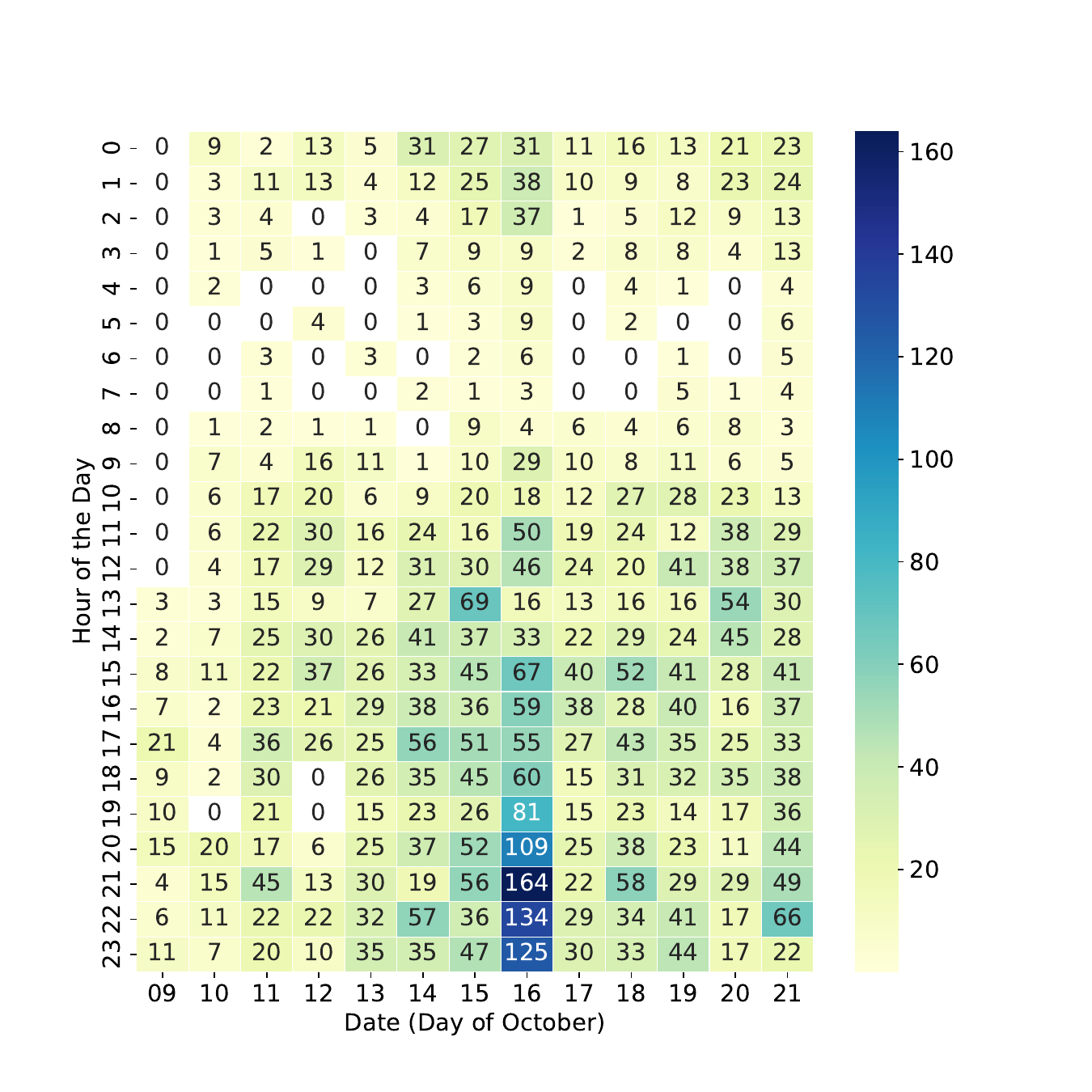}
  \caption{Queries submitted per hour (9-21 October). Midnight deadlines 16 October (Lab) and 21 October (Project).%The x-axis shows days and y-axis represents hours. The late hours of October 16th show high usage of the CodeHelp before the deadline.}
  }
  \label{fig:heatmap}
\end{figure}

Figure \ref{fig:heatmap} illustrates when, over the two week period of the study (9--21 October), students submitted queries to CodeHelp.  A total of 6,051 queries were made by 813 students.  The intense period of activity, leading up to midnight on October 16th, corresponds to the deadline for the lab activity. Notably, there was considerable usage of CodeHelp after October 16th, when student interactions with the tool were ungraded. 

%when students were no longer graded for their interactions with the tool.

%Figure \ref{fig:histograms} shows the distribution of the number of queries asked by students both before and after the lab deadline.  Use prior to the lab deadline was not graded, but students were expected to complete this in order to provide an informed response to the reflection question.  Use after the lab deadline corresponds to voluntary use of CodeHelp while working on the final course project.  The deadline for this project was midnight on October 21st.  

\begin{figure}
\centering
  \includegraphics[width=\columnwidth]{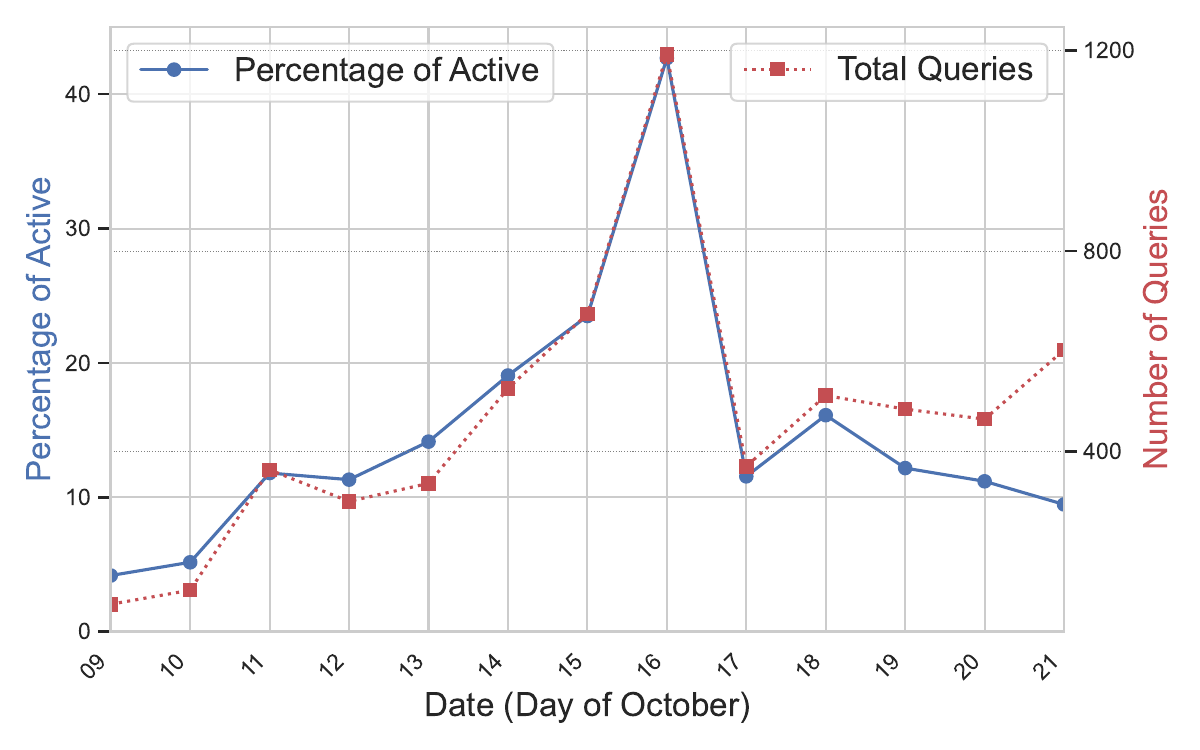}
  \caption{Percentage of students (out of all active) submitting queries per day and total number of queries.}
  \label{fig:histograms}
\end{figure}

Figure \ref{fig:histograms} shows the percentage of all active students using CodeHelp each day (left-hand axis), along with the total number of queries to CodeHelp (right-hand axis).  After the lab deadline (16 October), the percentage of students using CodeHelp remained fairly stable (although was trending downwards), but the number of daily queries increased as the project deadline drew nearer.  Given that this corresponds to voluntary use of CodeHelp, it suggests students saw value in the tool for supporting their coursework.

%the distribution of the number of queries asked by students both before and after the lab deadline.  Use prior to the lab deadline was not graded, but students were expected to complete this in order to provide an informed response to the reflection question.  Use after the lab deadline corresponds to voluntary use of CodeHelp while working on the final course project.  The deadline for this project was midnight on October 21st.  

% \begin{figure}
% \centering
%   \includegraphics[width=\columnwidth]{figs/Histograms_scaled.pdf}
%   \caption{Histogram of the number of queries, prior to and after the lab deadline.  Prior to the lab deadline, 562 students submitted a query, 483 submitted after the deadline (with 229 submitting in both periods).}
%   \label{fig:histograms}
% \end{figure}

\begin{figure*}
\centering
  \includegraphics[width=\textwidth]{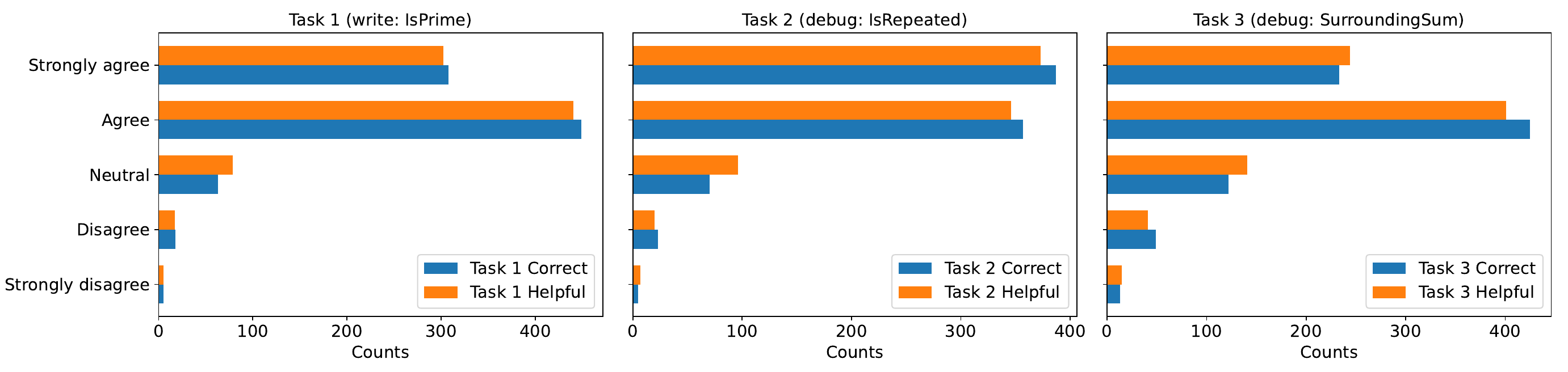}
  \caption{Student feedback on the correctness and helpfulness of responses from CodeHelp to the three lab questions.}
  \label{fig:student_feedback}
\end{figure*}

Figure \ref{fig:student_feedback} summarizes student responses to the Likert-scale items regarding the correctness and helpfulness of CodeHelp's responses to the three questions on the lab.  A large majority of the students either ``Strongly agree'' or ``Agree'' that the responses generated by CodeHelp were both correct and helpful.

\subsection{Desired characteristics}
%We asked students the following reflection question: \emph{``What do you think are the most important characteristics for an automated programming assistant (like LLM-TA) if it is to be suitable for use in a programming course?''} 

Our qualitative analysis of student responses on the important characteristics for a digital TA revealed three key themes: \textbf{Scaffolding} (i.e., \textit{how} student learning should be supported); \textbf{Appropriateness} (i.e., the \textit{nature} and \textit{content} of explanations and feedback); and \textbf{User Experience} (i.e., the \textit{values} that students express when considering their learning experience). Each theme highlights a crucial aspect of the educational support system students would expect from an AI Teaching Assistant.

%\subsubsection{Explanatory Feedback}
%The most prominent theme that emerged from the responses is the need for "Explanatory Feedback." Students indicated a strong preference for a programming assistant that provides detailed, step-by-step explanations of bugs, errors, or concepts without outright giving the code solutions. They value feedback that aids their understanding of the underlying principles behind coding tasks, which in turn fosters independent problem-solving. This theme underscores the educational role of the assistant, moving beyond error correction to enhance student knowledge.

\subsubsection{Scaffolding}

Students reported wanting to understand the content and felt that explanations that focused on process (\emph{how}) and foundational knowledge (\emph{why}) were key features.  They did not want to be directly provided with the solutions to the problems, but rather, they wanted to be taught how to obtain the solutions themselves (agency) to develop deep understanding. The student reflections included, among others, the following:

\begin{quote}
\emph{``One key idea is that it shouldn't supply code.''}
\end{quote}

\begin{quote}
\emph{``Therefore the assistant must make sure to be helpful without directly writing code or debugging code for the user by giving ideas and explaining logic and asking relevant questions to help the user understand.''} 
\end{quote}

%\begin{quote}
%\emph{"This includes understanding the problem, suggesting possible solutions, and guiding the user in implementing these solutions"}
%\end{quote}

\noindent Students also described the importance of support for focusing on relevant code sections:

\begin{quote}
\emph{``Like suggesting changing a small portion \ldots as people can focus on one section of the code.''}    
\end{quote}

% \begin{quote}
% \emph{``The information provided should be specific to the problem provided and not information based on general topics.''}
% \end{quote}

\noindent Some students emphasized the benefits of having support in problem-solving guidance:

\begin{quote}
\emph{``\ldots useful for understanding code or a complex algorithm as they break the problem down into simpler steps to understand.''}    
\end{quote}

%\noindent Others stressed the need to receive hints about how to progress:

\begin{quote}
\emph{``This may be done by asking the student questions instead of giving answers, such that the questions provided give the student a clue as to what to do.''}
\end{quote}

\noindent Several students talked about the value of multiple exemplars which is a good pedagogical practice:

\begin{quote}\emph{``\ldots beneficial to have an assistant that presents a few different ways of solving an issue, to make students reflect more, but also to make sure there is a solution that could be implemented with varying levels of coding experience.''}
\end{quote}

\begin{quote}
\emph{``The ability to provide explanations on how to write code that fulfills the specifications in various forms so that the user can expand their coding repertoire. ''}
\end{quote}

\noindent Finally, they also talked about breaking up explanations in ways that may reduce cognitive effort:

\begin{quote}
\emph{``For writing of code I think the AI should break it down into simple steps for the student to focus on and write rather than writing the code for the student, as this will help the student learn.'' 
%``\ldots thoroughness in explanation and ability to break down larger problem into smaller, manageable steps'' 
}
\end{quote}

\begin{quote}
\emph{``The tool must also provide good explanations instead of just solutions and should be able to break down unfamiliar concepts in simple terms.''}
\end{quote}
\subsubsection{Appropriateness}
The second theme that emerged is the need for the presented concepts and explanations as well as language expressions to be appropriate for the level of the audience. 

\begin{quote}
\emph{``\ldots [an] important characteristic would be understandability. If the feedback provided was not written in a well-structured manner, this would require a lot of effort for us to slowly understand and read\ldots''}
\end{quote}
%, wasting perhaps even more time than required to analyze the problem ourselves.''}

\begin{quote}
\emph{``\ldots It should be detailed enough that the person asking will be able to understand the information provided and how to fix a certain issue \ldots''}
\end{quote}

\begin{quote}
\emph{``Not everyone will be at the same level of programming as others. Hence, the explanations should be customised for everyone so that users of different levels of coding experience can find this assistant useful.'' %This could mean not overusing jargon when not required etc.''
}
\end{quote}

\noindent Many students clearly preferred explanations in plain language, avoiding complex jargon:

\begin{quote}
\noindent\emph{``conciseness is important\ldots The information provided [by the model] should be specific to the problem provided and not information based on general topics.'' }
\end{quote}

%\begin{quote}
%\emph{``An automated programming assistant like CodeHelp should be accurate, easy to use, and provide clear and concise explanations to the user.'' }
%\end{quote}

\begin{quote}
\emph{``I think the most important characteristic is providing help in a way that is easier to understand. This means organising hints in a bulleted list, using simple language\ldots''} 
\end{quote}

\noindent Further, the content should be constrained to the topics that have been covered in the course, and which are appropriate for beginners.

\begin{quote}
\emph{``and for more complex problems, providing snippets of code that beginners may not be familiar with yet.''}
\end{quote}

\begin{quote}
\emph{``\ldots showing solutions using only methods taught in class\ldots. ChatGPT for example fails at this often\ldots [it] may suggest far more complex solutions that a beginner cannot yet understand.''}
\end{quote}

%The solutions also have to be explained at an appropriate level for beginners, which CodeHelp did well.

\subsubsection{User Experience}
The final theme encompasses the values that students describe as important for their experience when using the tool for learning.  They emphasise the importance of agency:

\begin{quote}
\emph{``broad cues as help could be more effective in letting students `find' their own solutions from these cues.''}
\end{quote}

\begin{quote}
\emph{``help the student figure out their own way of solving the problem.''}
\end{quote}

 \begin{quote}
\emph{``It should also be interactive, allowing students to try things for themselves whilst providing feedback and error analysis.''}
\end{quote}

\noindent Students wanted rapid responses with support always available:

\begin{quote}
\emph{`` Sometimes you can't always get help or assistance from someone who is experienced in coding and can help you, so to have a 24/7 assistance that can achieve the same level as a tutor/teacher is really helpful.''}
\end{quote}

\begin{quote}
\emph{``helps explain codes without the help of a tutor, helps the student to get help all the time everywhere.''}
\end{quote}

\noindent Most importantly, the responses from the tool must be correct.   Trustworthiness and precision in the feedback are crucial for students to be confident in using such tools for educational purposes:

\begin{quote}
\emph{``I think the accuracy is very important. Having an automated programming assistant that is very accurate is essential.''}
\end{quote}

In summary, the thematic analysis indicates that students seek to enhance their learning by ensuring they understand the content and develop their critical thinking and problem-solving skills. They want an educational aid rather than a solution provider. They desire a tool that helps them understand their coding errors and guides them to correct those mistakes, all communicated in clear and concise language and with accurate feedback that they can trust.

\section{Discussion}

The use of AI Teaching Assistants (TAs) is likely to increase, especially given their cost effectiveness relative to human TAs. In this study, we used the most expensive model from OpenAI at the time (GPT-4). The total cost for handling 
%1045 
6,051
queries spread across all hours of the day over 2 weeks was US\$510, roughly equivalent to a little over 30 hours of human TA time (assuming a typical rate of US\$15 per hour). Such tools may not entirely replace human TAs, and we do not advocate for that, but they may be a useful way to augment human TA support, especially by answering common queries outside of regular working hours---a time when our prototype experienced substantial student usage. Usage spiked dramatically leading up to the deadline, 
with 20\% of all queries being submitted on the 16th October, and many of those in the hours immediately before the lab deadline. 
%with 46.2\% of queries received after the deadline. 
Future research could investigate the impact of always available support on student procrastination, given previous findings linking automatic grading to increased procrastination~\cite{auvinen2015harmful}. 
%fouh2014exploring

%Our analysis, based on responses from over 700 students, 

Our qualitative results align with findings from previous research, highlighting a trend where students weigh the convenience of help resources against perceived learning benefits~\cite{hou2024effects}. Notably, students preferred shorter responses that facilitated their problem-solving process in a structured manner rather than receiving solutions directly. This also aligns with students' responses about in-person TAs~\cite{lim2023student}, however this same prior work found that students sometimes want direct answers when seeking help close to a deadline.  Future work could examine if student opinions change (or if they would opt for just a solution) when near a deadline. It's also possible that students felt more comfortable expressing maladaptive learning behaviors (just wanting solutions) in the 1-1 interviews conducted by \citeauthor{lim2023student} than in our questionnaire. %than in a feedback form as done here. 

Encouragingly, this explicit desire for solutions to remain hidden suggests that students want to use AI TAs for learning, not to obtain solutions. This contrasts with the concerns expressed by academics about use of AI by students to cheat~\cite{prather2023robots, zastudil2023generative, sheard2024instructor}. Prior work provides evidence that students are most likely to cheat when they run out of time, when the workload is too high, and when assignments are too hard~\cite{sheard2003determination}. AI TAs may provide a higher level of individual support to students that may reduce time spent waiting for feedback, hints, and other teaching support humans provide. A recent study also highlighted the on-demand nature of help from AI tools as a primary motivator for student use~\cite{hou2024effects}. AI TAs that provide immediate and personalized feedback may therefore \textit{reduce} academic misconduct as students can get assistance promptly and overcome some of the pressures that lead to cheating behaviours. 

However, it is crucial for instructors to continue to emphasize the purpose and value of learning activities and assignments. In a recent interview study, some students claimed they would resort to using AI tools to complete `busy work', referring to activities perceived as lacking educational value~\cite{zastudil2023generative}. So, while clearly the design of AI TAs is critical, instructors also play a pivotal role in ensuring ethical use of AI TAs in the learning process.

\subsection{Design Recommendations} 

%Based on our qualitative analysis, we propose the following design recommendations. 
%\subsubsection{Preserve Students' Agency} 
%Students expressed the desire to independently arrive at a solution rather than being provided with solutions directly. This preference aligns with recent research outside of computing education which demonstrated that users are more likely to engage with AI tools when they feel a sense of ownership over the process and outcomes~\cite{oh2018lead, huang2023causalmapper, yuan2022wordcraft}. This is a promising area for future research to explore the impact of AI Teaching Assistants on students' agency in their learning process.  

\subsubsection{Scaffold a Learning Process that Preserves Students' Agency} 

Students expressed the desire to independently arrive at a solution rather than being provided with solutions directly. This preference aligns with recent research outside of computing education which demonstrated that users are more likely to engage with AI tools when they feel a sense of ownership over the process and outcomes~\cite{oh2018lead, huang2023causalmapper, yuan2022wordcraft}. 
Exploring the impact of AI Teaching Assistants on students' agency in their learning process would be a promising area for future research. While students want their agency to be preserved, they still appreciated hints and guidance in the form of procedural scaffolding. Rather than receiving direct instructions about how to fix their code, students wanted to have their attention directed to the area within the code that needed attention or help with understanding the problem they are trying to solve. 

\subsubsection{Tailor Scaffolding and Responses to Students' Expertise}

Students wanted responses to be aligned with their knowledge and abilities. Students highlighted issues with current tools like ChatGPT, which fail to account for their expertise level, resulting in confusing and unhelpful programming patterns. However, they also expressed interest in being exposed to new patterns and examples that could challenge and expand their abilities. AI Teaching Assistants should therefore maintain user models that can be used to guide process and personalize responses. In addition, students advocated for responses to be clearly and concisely presented, such as by using bullet points. This aligns with best practice in feedback design~\cite{krause2017critique}. These findings underscore the importance of providing well-organized responses that are easily comprehensible.

%Processing feedback is a cognitively demanding task that requires the feedback receiver to read, interpret, and prioritize the feedback they receive~\cite{oleson2020computing}. Notably, students in our study advocated for breaking down responses into smaller components, aligning with the existing preference for short and specific feedback~\cite{krause2017critique}. These findings reinforce the importance of delivering responses that are easily digestible and manageable for effective learning outcomes.

%% As a replacement for a TA, a digital TA has to be correct in a way that other AI generated resorces don't need to be (unlike ICER paper where some value in seeing variety of queality

\section{Conclusion}
%In this study, we explored key implications arising from students' reflections about using AI-assisted learning tools in the domain of computing education. We find that students used an AI TA to help with their assignments and that they self-report valuing the quality and accuracy of the AI TA responses, and that the responses were tailored toward helping students learn the material rather than simply giving away the answer.

In this study, we explored key implications arising from students' reflections about using AI-assisted learning tools in the domain of computing education. Our findings underscore two key points. First, students not only enjoyed using the AI TA but also engaged with it extensively, especially late at night leading up to the lab deadline. Second, students desired more than mere answers; seeking agency in their learning process and emphasizing the need for scaffolding to guide them through the problem-solving process, and to help them generate their own solutions.

\begin{acks}
The authors would like to thank Mark Liffiton for developing the Gen-Ed framework and CodeHelp, and to both Mark Liffiton and Brad Sheese for their contributions to prior research on CodeHelp.
\end{acks}

%while aligning with their individual path.

% Steve: 
%In this study, we explored key implications arising from students' reflections about using AI-assisted learning tools in the domain of computing education. Our findings underscore to key points. First, students not only enjoyed using the AI TA but also engaged with it extensively, especially late at night leading up to the deadline. Second, students desired more than mere answers; seeking agency in their learning process and emphasizing the need for scaffolding to guide them toward the solution while aligning with their individual path. 

\balance
\bibliographystyle{ACM-Reference-Format}
\bibliography{iticse}

\end{document}